\documentclass[]{spie}  

 \usepackage{float}
 \usepackage{adjustbox}
\usepackage{amsmath,amsfonts,amssymb}
\usepackage{graphicx}
\usepackage[colorlinks=true, allcolors=blue]{hyperref}
\usepackage[colorinlistoftodos]{todonotes}
\usepackage{subcaption}
\usepackage{multirow}
\usepackage{algorithm}
\usepackage{algpseudocode}
\usepackage{booktabs}
\usepackage{amsmath}
\usepackage{graphicx,subcaption}
\usepackage{tikz}      
\usetikzlibrary{arrows.meta, calc, decorations.markings, positioning}
\usepackage{siunitx} 
\usepackage{array}

\title{A comparison of CNN architectures for Alzheimer's disease detection in single-view MRI scans}

\author[a]{Hiram Zuniga}
\author[a,$\ast$]{Ulises Orozco-Rosas}
\author[a]{Kenia Picos}
\affil[a]{CETYS Universidad, Ave. CETYS Universidad No. 4, Fracc. El Lago, C.P. 22210, Tijuana, Baja California, Mexico}

\authorinfo{*Corresponding author(s).\\U. Orozco-Rosas: E-mail: ulises.orozco@cetys.mx\\[4pt]
Author's version of a manuscript accepted for presentation at SPIE Optics
+ Photonics 2026. Copyright 2026 SPIE.}

\begin{document} 
\maketitle


\begin{abstract}

Alzheimer's disease is a leading cause of death with no cure. Therefore, early detection is critical to slow progression and preserve quality of life. Diagnosis relies on medical history, cognitive tests, physical exams, and MRI brain scans, making deep learning suitable for Alzheimer's classification. This work proposes a benchmark that evaluates ten different convolutional neural network (CNN) architectures (including ResNet, DenseNet, MobileNet, EfficientNet, and VGG family models) under the same held-out test split protocol. A two-stage transfer learning and full fine-tuning pipeline is introduced to perform training using a class-balanced subset (3,900 images) derived from the OASIS medical imaging dataset, comprising 86,437 single-view MRI brain scans labeled into four classifications of Alzheimer's disease: Non-Demented, Very Mild Dementia, Mild Dementia, and Moderate Dementia. The best results were achieved by VGG16, with a 0.9637 validation accuracy and a 0.9533 test accuracy score. A key finding documented in this work is the difficulty of classifying the transition from Non-Demented to Very Mild Demented stages, observed consistently across all ten architectures.

\end{abstract}




\section{Introduction}

Alzheimer's disease (AD) has been the seventh leading cause of death in the United States in recent years~\cite{11}, and is the most common cause of dementia in the elderly~\cite{2}, affecting nearly one-half of Americans older than 85 years old~\cite{Winslow2011TreatmentAlzheimer} and 22\% of all persons aged 50 and above~\cite{10}. AD impacts the lives of nearly 7.2 million people older than 65 in the US, and this number could grow to 13.8 million in 2060 (almost double).  AD has caused nearly 120,122 deaths in the US, only in 2022. Between 2000 and 2022, the reported deaths from AD have increased by more than 142\%, so not only is AD not going anywhere but is increasing the number of deaths and the number of people affected at a rapid rate~\cite{11}.

Despite advances in Alzheimer's research, there is still no current treatment to cure the disease, no clearly effective disease-modifying treatments~\cite{Petrella2013NeuroimagingCureAD,Rasmussen2019WhyEarlyDiagnosis}, and no pharmacologic agents can reverse the progression~\cite{Winslow2011TreatmentAlzheimer,Rasmussen2019WhyEarlyDiagnosis}. Alzheimer's disease is a fatal degenerative dementing disorder, which starts with memory deterioration that progresses to debilitating loss of mental and physical faculties~\cite{1,3}. Early detection of this disease have been seen to be relevant, because you can treat reversible causes of dementia~\cite{8} and implement strategies to reduce or prevent further progression of the disease's, improving quality of life~\cite{Rasmussen2019WhyEarlyDiagnosis,8}, some of the treatments and recent medicine advances has shown that donepezil is effective in stabilizing or slowing progressive decline in cognition, function, and behavior~\cite{4}.

Built upon these ideas, early diagnosis is important because it can represent a big clinical advantage in initiating treatment early in the course of the disease~\cite{4}, and it also gives the patient time to make choices or plan ahead for the future and their carers~\cite{Rasmussen2019WhyEarlyDiagnosis}. Currently, AD diagnosis has been dependent mainly on clinical suspicion based on the patient’s or caregiver concerns~\cite{8}, medical history~\cite{12}, psychometric scales, Neuropsychological testing to evaluate patient cognitive function, and neuroimaging as an auxiliary diagnostic tool; these include magnetic resonance imaging (MRI) to detect brain structural changes~\cite{9}. This makes deep learning and medical imaging a suitable aid for diagnosis using these MRI brain scans. 

Deep learning techniques have been widely explored for Alzheimer classification using MRI brain scans \cite{Shaikh2025,Nagarajan2025,Raj2025}. This work presents a benchmark that evaluates ten different Convolutional Neural Network (CNN) architectures under the same protocol. To evaluate these neural networks, this study introduces a two-stage transfer learning and full-tuning pipeline: 5 epochs are used to train the head classifier of the CNN with a frozen backbone, and the remaining 20 epochs are used for full fine-tuning on the CNN architecture with an unfrozen backbone. 

\begin{figure}[t]
\centering
\includegraphics[width=\linewidth]{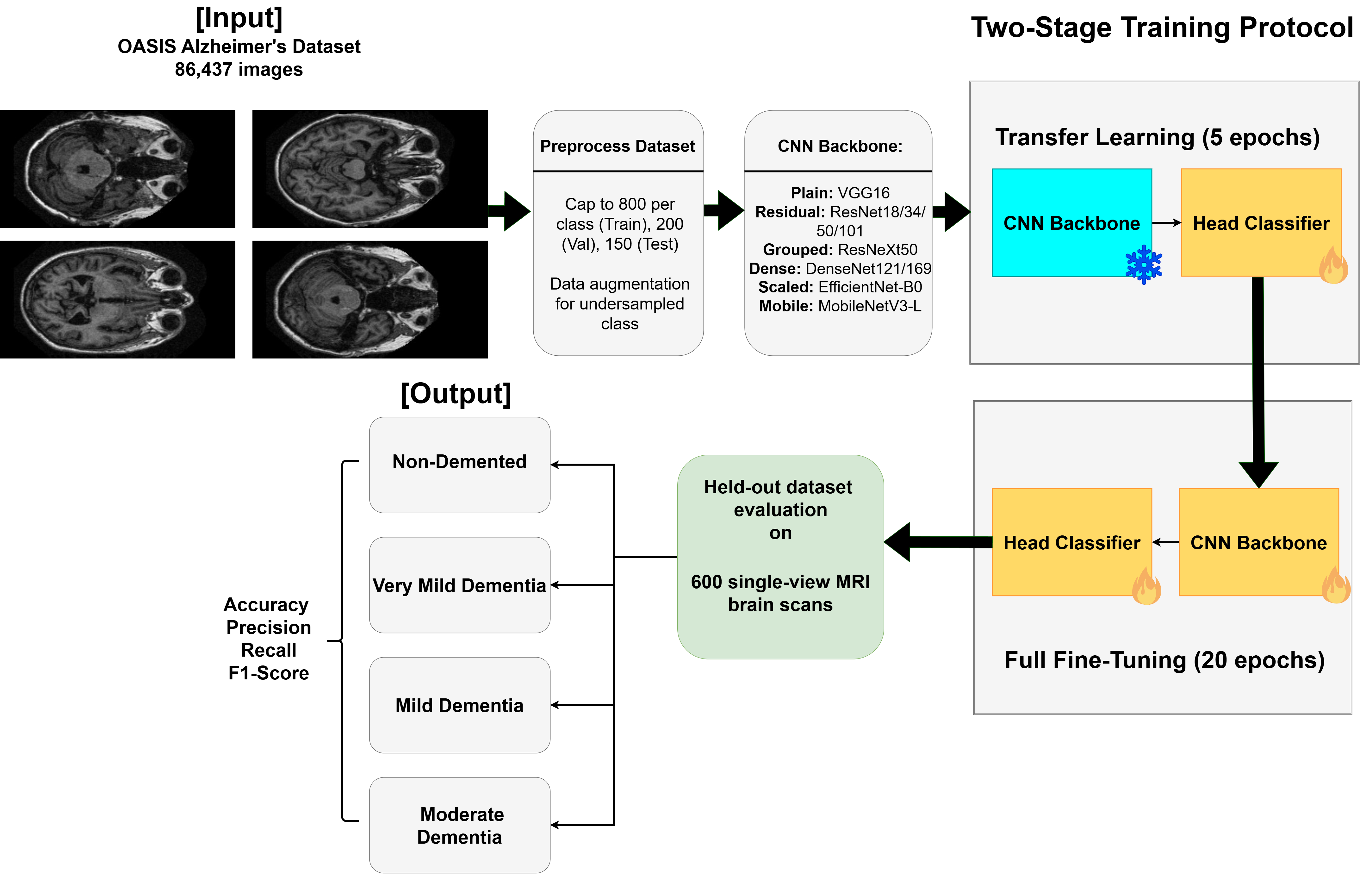}
\caption{Overview of the benchmark pipeline using the OASIS dataset proposing a two-stage training protocol. Snowflake and flame icons denote frozen and trainable
components, respectively.}
\label{fig:pipeline}
\end{figure}

The dataset used in this study is the OASIS medical dataset~\cite{Marcus2007OASIS}, comprising 86,437 single-view MRI brain scan images, labeled into four classes: Mild Dementia, Moderate Dementia, Very Mild Dementia, and Non Demented. During training across these different CNN architectures, a key finding was noted: all models struggle to differentiate between the Non Demented and the Very Mild Dementia classes. Additionally, an overview of the described pipeline is presented in Fig.~\ref{fig:pipeline}.

The key contributions of this work are described as follows:
\begin{itemize}
    \item A benchmark that evaluates ten different CNN architectures under the same evaluation protocol, comprising: the same split, the same image preprocessing, the same augmentation, sampler, batch size, and epochs. 
    \item A two-stage transfer learning and full fine-tuning pipeline, using 5 epochs for transfer learning with the frozen backbone and 20 epochs for fine-tuning with an unfrozen backbone. 
\end{itemize}

The remainder of the paper is organized as follows: Section \ref{sec:methodology} provides an overview of the materials and methods, describing the dataset, preprocessing, the architectures, and so on. Section \ref{sec:results} presents the results from running the experiments and provides a technical discussion on them. Section \ref{sec:conclusions} introduces the conclusion and discusses future research directions.


\section{Experimental Methodology}
\label{sec:methodology}
This section presents all of the methods used to obtain the results. Additionally, it introduces the dataset and the split, the CNN architectures benchmarked in the study, the evaluation metrics, and the experimental setup for reproducibility.

\subsection{Dataset and partitioning}
\label{subsec:dataset}
\begin{figure}[H]
\centering
\setlength{\tabcolsep}{0pt}%
\begin{subfigure}{0.40\textwidth}
  \includegraphics[width=\linewidth]{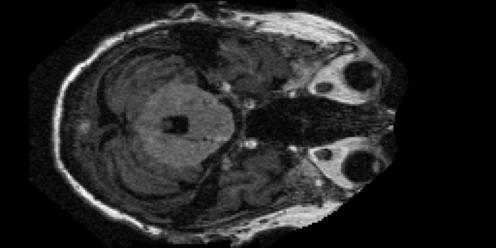}
  \caption{Mild}
\end{subfigure}\hspace{4pt}%
\begin{subfigure}{0.40\textwidth}
  \includegraphics[width=\linewidth]{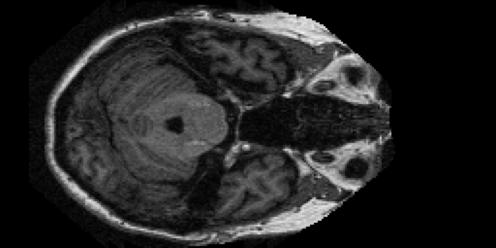}
  \caption{Moderate}
\end{subfigure}

\vspace{4pt}

\begin{subfigure}{0.40\textwidth}
  \includegraphics[width=\linewidth]{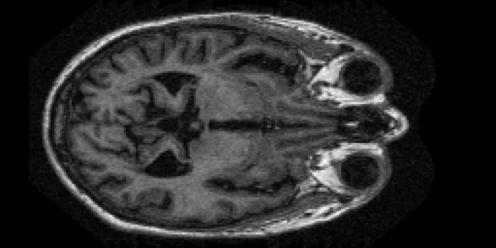}
  \caption{Non-Demented}
\end{subfigure}\hspace{4pt}%
\begin{subfigure}{0.40\textwidth}
  \includegraphics[width=\linewidth]{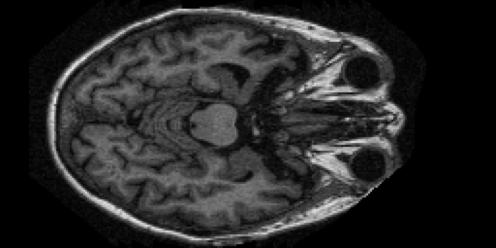}
  \caption{Very mild}
\end{subfigure}
\caption{Representative single-view MRI slices from the four OASIS classes: (a) Mild Dementia, (b) Moderate Dementia, (c) Non-Demented, and
(d) Very Mild Dementia.}
\label{fig:samples}
\end{figure}

This work uses the OASIS Alzheimer detection dataset \cite{Marcus2007OASIS} for training and evaluation, having 86,437 single-view MRI brain scan images. This dataset is based on 416 subjects aged 18 to 96 years, and images are divided into 4 classes: Mild Dementia with 5,002 images, Moderate Dementia with 488 images, Very Mild Dementia with 13,725 images, and Non Demented with 67,222 images. Based on the number of images per class, a class imbalance problem is notable, more specifically for the Moderate Dementia class, which has only 488 images compared to the 67,222 images in the Non-Demented class. Dataset details are shown in Table \ref{tab:oasis-full}. Additionally, some samples from the dataset are shown in Fig.~\ref{fig:samples}.

\begin{table}[H]
\caption{Class distribution of the OASIS single-view MRI dataset.}
\label{tab:oasis-full}
\begin{center}
\begin{tabular}{lrr}
\hline
\rule[-1ex]{0pt}{3.5ex} Class & Images & Share (\%) \\
\hline
\rule[-1ex]{0pt}{3.5ex} Mild Dementia      &  5{,}002 &  5.79 \\
\rule[-1ex]{0pt}{3.5ex} Moderate Dementia  &    488 &  0.56 \\
\rule[-1ex]{0pt}{3.5ex} Non-Demented       & 67{,}222 & 77.77 \\
\rule[-1ex]{0pt}{3.5ex} Very Mild Dementia & 13{,}725 & 15.88 \\
\hline
\rule[-1ex]{0pt}{3.5ex} \textbf{Total}     & 86{,}437 & 100.00 \\
\hline
\end{tabular}
\end{center}
\end{table}

To mitigate the class imbalance effect, this study caps all images to 800 for the training cohort, 200 for validation, and 150 for a held-out test split. The reason to cap the images from the dataset instead of applying techniques like class weighting, random undersampling (RUS), synthetic minority over-sampling technique (SMOTE), and others is that the key objective of this work is to benchmark these ten different architectures under single evaluation conditions, rather than maximizing the accuracy value that can be achieved. 

Since the natural cap is 488 for the Moderate Dementia class, this study divides these images into 100 for training, 200 for validation, and 150 for testing. The reason is to have original images for the validation and testing splits. Additionally, augmentation is applied to training samples to reach 800 samples for the other classes. Augmented samples are deliberately excluded from the validation and test splits, as the study requires a clean and consistent comparison across architectures. The details on how many images are available during the experiments are available in Table \ref{tab:dataset_classes}.

\begin{table}[H]
\caption{Class-balanced partition of the OASIS single-view MRI dataset used in this work. The training column reports the number of distinct original images; the balanced sampler described in Section~\ref{sec:preprocessing} fixes the number of samples drawn per epoch at 800 per class.}
\label{tab:dataset_classes}
\begin{center}
\begin{tabular}{lrrrr}
\hline
\rule[-1ex]{0pt}{3.5ex} Class & Train (original) & Train (per epoch) & Validation & Test \\
\hline
\rule[-1ex]{0pt}{3.5ex} Mild Dementia       &   800 &   800 & 200 & 150 \\
\rule[-1ex]{0pt}{3.5ex} Moderate Dementia   &   100 &   800 & 200 & 150 \\
\rule[-1ex]{0pt}{3.5ex} Non-Demented        &   800 &   800 & 200 & 150 \\
\rule[-1ex]{0pt}{3.5ex} Very mild Dementia  &   800 &   800 & 200 & 150 \\
\hline
\rule[-1ex]{0pt}{3.5ex} \textbf{Total}      & 2,500 & 3,200 & 800 & 600 \\
\hline
\end{tabular}
\end{center}
\end{table}

\subsection{Preprocessing and class balancing}
\label{sec:preprocessing}
To ensure a consistent evaluation, the same preprocessing pipeline was applied across the 10 CNNs. First, the images are preprocessed to a single grayscale channel; they are resized to 224$\times$224 pixels, converted to tensors, and normalized with a mean value of 0.5 and a standard deviation of 0.5. This normalization was used instead of ImageNet standard normalization since RGB does not apply in this setting, and only one Grayscale channel is used. 

A Weighted Random Sampler is introduced with an objective of 3,200 samples, meaning that the Moderate Dementia class is sorted 8 times per epoch, and in each sort, augmentations are applied consisting of random affine with degrees = 10 and color jitter with a brightness value of 0.08 and a contrast of 0.08. All of these new samples are generated on the fly, rather than precomputed and stored beforehand. This enables matching the same 800 images for training the other classes. An important distinction is that augmentation is not applied to the validation and testing splits. This ensures that the evaluation reflects original OASIS data, which is closer to what the model would encounter at inference, instead of artificially altered images. 

ImageNet weights expect a three-channel RGB input. Since this classification is performed on single-channel grayscale MRI brain scans, most of the pretrained architectures are not compatible with this input. This work replaces the first convolutional layer with an equivalent layer expecting one input channel instead of three, and its weights are initialized by averaging the pretrained weights across the three RGB channels, as expressed in Eq.~\eqref{eq:graystem}, where $\mathbf{W}^{(1)}_{c}$ denotes the pretrained weights of the first convolutional layer corresponding to channel $c$.

\begin{equation}
\mathbf{W}^{(1)}_{\text{gray}} = \frac{1}{3}\sum_{c=1}^{3}\mathbf{W}^{(1)}_{c}
\label{eq:graystem}
\end{equation}

This method retains features learned during ImageNet pretraining, rather than reinitializing the first layer with random weights. This procedure was applied uniformly across the ten architectures, adapting only the location where the first convolutional layer resides in each model, as it is defined differently across architectures.

\subsection{CNN architectures under comparison}
\label{subsec:CNNs}
Ten different CNN architectures are evaluated to compare performance in the Alzheimer classification task. These different CNNs were selected to cover different design families such as plain stack, residual, grouped residual, densely connected, compound scaled, and mobile-oriented architectures~\cite{9490209, Lopez-Montiel2025}. This inclusion of heterogeneous architectures provides a complete comparison across different CNNs with different purposes. All of these CNNs were initialized with IMAGENET1K\_V1 weights from torchvision, pretrained on ImageNet\cite{Deng2009ImageNet}, and are introduced in Table \ref{tab:architectures}.

\begin{table}[H]
\caption{CNN architectures evaluated in this work. Head parameter counts correspond to the classification head only, and total parameter counts include the single-channel first convolutional layer.}
\label{tab:architectures}
\begin{center}
\renewcommand{\arraystretch}{1.3}%
\begin{tabular}{llrrc}
\hline
\rule[-1ex]{0pt}{3.5ex} Architecture & Design family & Total params & Head params & Pretrained head \\
\hline
\rule[-1ex]{0pt}{3.5ex} VGG16~\cite{Simonyan2015VGG}            & Plain deep stack  & 134.3\,M & 119,562,244 & Yes \\
\rule[-1ex]{0pt}{3.5ex} ResNet18~\cite{He2016ResNet}            & Residual          &  11.3\,M &     132,356 & No  \\
\rule[-1ex]{0pt}{3.5ex} ResNet34~\cite{He2016ResNet}            & Residual          &  21.4\,M &     132,356 & No  \\
\rule[-1ex]{0pt}{3.5ex} ResNet50~\cite{He2016ResNet}            & Residual          &  24.0\,M &     525,572 & No  \\
\rule[-1ex]{0pt}{3.5ex} ResNet101~\cite{He2016ResNet}           & Residual          &  43.0\,M &     525,572 & No  \\
\rule[-1ex]{0pt}{3.5ex} ResNeXt50-32x4d~\cite{Xie2017ResNeXt}   & Grouped residual  &  23.5\,M &     525,572 & No  \\
\rule[-1ex]{0pt}{3.5ex} DenseNet121~\cite{Huang2017DenseNet}    & Densely connected &   7.2\,M &     263,428 & No  \\
\rule[-1ex]{0pt}{3.5ex} DenseNet169~\cite{Huang2017DenseNet}    & Densely connected &  12.9\,M &     427,268 & No  \\
\rule[-1ex]{0pt}{3.5ex} EfficientNet-B0~\cite{Tan2019EfficientNet} & Compound-scaled   &   4.3\,M &     328,964 & No  \\
\rule[-1ex]{0pt}{3.5ex} MobileNetV3-L~\cite{Howard2019MobileNetV3} & Mobile / NAS      &   4.2\,M &   1,235,204 & No  \\
\hline
\end{tabular}
\end{center}
\end{table}
As shown in Table \ref{tab:architectures}, the evaluated architectures span a wide range of model sizes, from 4.2M parameters for the mobile and efficient CNNs to 24M parameters for the larger ResNet models. Additionally, ResNet101 and VGG16 have 43M and 134.3M parameters, respectively; these architectures represent larger models, and this benchmark will let us evaluate the difference in computational time and performance between larger and smaller models to assess whether large architectures pose an advantage in this Alzheimer classification task.

\begin{table}[H]
\caption{Shared classification head applied to eight of the ten architectures.
The input dimension $d$ is the feature dimension of each backbone
(e.g., 512 for ResNet18/34, 2,048 for ResNet50/101 and ResNeXt50).
VGG16 and MobileNetV3-L use different heads, as described in the text.}
\label{tab:head}
\begin{center}
\renewcommand{\arraystretch}{1.3}%
\begin{tabular}{clc}
\hline
\rule[-1ex]{0pt}{3.5ex} Layer & Type & Output dim. \\
\hline
\rule[-1ex]{0pt}{3.5ex} 1 & Linear ($d \rightarrow 256$)      & 256 \\
\rule[-1ex]{0pt}{3.5ex} 2 & ReLU                              & 256 \\
\rule[-1ex]{0pt}{3.5ex} 3 & Dropout ($p=0.5$)                 & 256 \\
\rule[-1ex]{0pt}{3.5ex} 4 & Linear ($256 \rightarrow 4$)      & 4   \\
\hline
\end{tabular}
\end{center}
\end{table} 

Eight of the ten architectures receive the same classification head, described in Table \ref{tab:head}. Two exceptions apply. VGG16 retains its original pretrained classifier, consisting of two layers of 4096 neurons and only replacing the output layer with one compatible with four classes. This classifier design was kept as it represents 89\% of the whole model (Table \ref{tab:architectures}) and replacing it with the classifier described in Table \ref{tab:head} would have discarded the majority of the pretrained model. MobileNetV3-L follows the same structure as the shared head but preserves a 1280-unit hidden layer of its original design.

\subsection{Two-stage transfer-learning and fine-tuning protocol}
\begin{figure}[H]
\centering
\includegraphics[width=\linewidth]{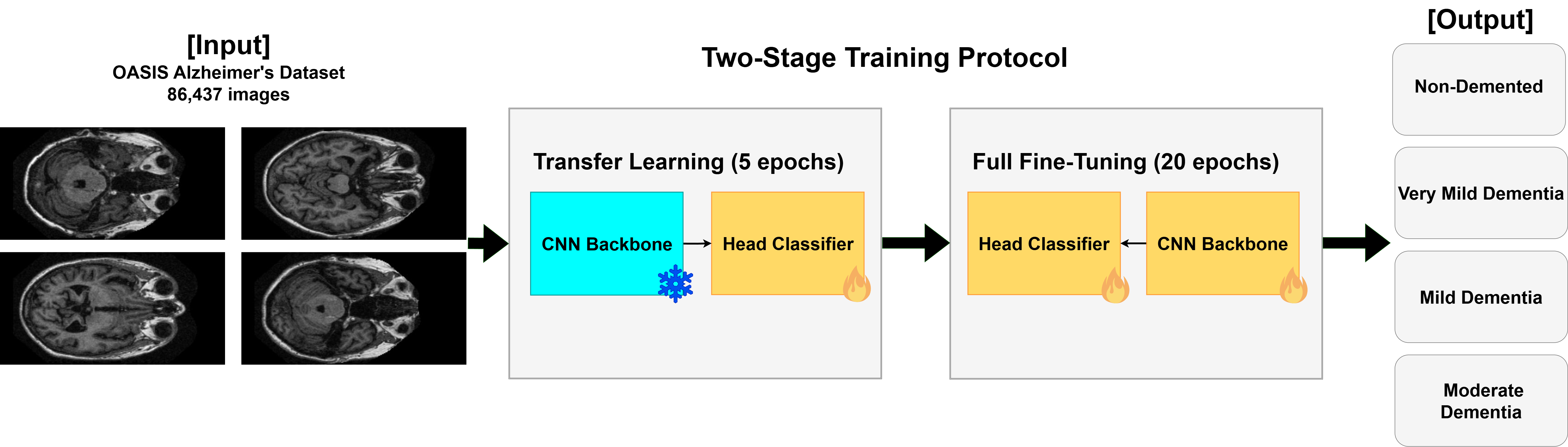}
\caption{Two-stage transfer-learning and fine-tuning protocol, applied
identically to all ten architectures. Snowflake and flame icons
denote frozen and trainable components, respectively.}
\label{fig:twostage}
\end{figure}

This work introduces a two-stage transfer-learning and fine-tuning pipeline for training, applied identically to the ten models to ensure consistent training conditions across all architectures. The first phase is transfer learning: training only the head classifier while keeping the encoder backbone frozen, with a learning rate of $1\times10^{-4}$ for 5 epochs, as this phase is a head warm-up before adapting the full architecture. Phase two is full fine-tuning; after the first 5 epochs, the backbone is unfrozen all at once, and then 20 epochs are applied to train the full encoder with a $1\times10^{-5}$ learning rate, which is lower to reduce the risk of overfitting, since this risk is higher when adapting the full architecture rather than only the head. Both phases use the Adam optimizer and are trained on a GPU. A representative diagram of this two-stage pipeline is shown in Fig~\ref{fig:twostage}.

Training in two stages avoids the risk of overfitting that arises when full fine-tuning is applied from the first epoch. This is especially true with the small number of images used for training (800 images) and the large number of parameters and layers some architectures can reach. Therefore, phase 1 stabilizes the head classifier before entering full fine-tuning within the MRI domain. 

\subsection{Evaluation protocol and metrics}
To ensure a consistent comparison between architectures, this work introduces an identical evaluation protocol for CNNs. In preprocessing, the same channel adaptation was used for all ten models. During training, the same sampler, batch size, optimizer, loss function, number of epochs, and learning rate were applied across all ten CNNs. Additionally, the same held-out test split with the same seed was used for evaluation across all ten models. The only thing that changed across these ten architectures was the backbone and that is what makes this evaluation work, and the head exception already declared in the subsection \ref{subsec:CNNs}. More details on these training configurations are available in Table \ref{tab:hyperparams}.  

\begin{table}[H]
\caption{Training configuration, identical for all evaluated architectures.}
\label{tab:hyperparams}
\begin{center}
\begin{tabular}{ll}
\hline
\rule[-1ex]{0pt}{3.5ex} Parameter & Value \\
\hline
\rule[-1ex]{0pt}{3.5ex} Input resolution        & $224 \times 224$, single channel \\
\rule[-1ex]{0pt}{3.5ex} Normalization           & mean $0.5$, std $0.5$ \\
\rule[-1ex]{0pt}{3.5ex} Batch size              & 4 \\
\rule[-1ex]{0pt}{3.5ex} Samples per epoch       & 3,200 (800 per class) \\
\rule[-1ex]{0pt}{3.5ex} Phase 1 -- epochs       & 5 (classification head only) \\
\rule[-1ex]{0pt}{3.5ex} Phase 1 -- optimizer    & Adam, learning rate $1\times10^{-4}$ \\
\rule[-1ex]{0pt}{3.5ex} Phase 2 -- epochs       & 20 (all parameters) \\
\rule[-1ex]{0pt}{3.5ex} Phase 2 -- optimizer    & Adam, learning rate $1\times10^{-5}$ \\
\rule[-1ex]{0pt}{3.5ex} Loss function           & Cross-entropy \\
\rule[-1ex]{0pt}{3.5ex} Model selection         & Highest validation accuracy \\
\rule[-1ex]{0pt}{3.5ex} Random seed             & 42 \\
\hline
\end{tabular}
\end{center}
\end{table}

This work evaluates the models on their best epoch. A checkpoint retains the best-performing parameters on the validation split during training across the 25 epochs. Additionally, the held-out test split was used only for final evaluation; it was not used during training or validation to ensure no data leakage occurred before the final test. This held-out test was only used to report final metrics, and the results did not influence training or reporting decisions to ensure the benchmark holds value.  

Metrics reported in the benchmark are global accuracy, the macro- and weighted-average versions of precision, recall, and F1-score, and support, all described in Eq.~\eqref{eq:acc}--\eqref{eq:support}. For a given class~$k$, $\mathrm{TP}_k$,
$\mathrm{FP}_k$, and $\mathrm{FN}_k$ denote true positives, false positives,
and false negatives; $K=4$ is the number of classes; $n_k$ is the support of
class~$k$; and $N=\sum_{k} n_k$ is the total number of test samples.

\begin{equation}
\mathrm{Accuracy} = \frac{\sum_{k=1}^{K}\mathrm{TP}_k}{N}
\label{eq:acc}
\end{equation}

\begin{equation}
P_k = \frac{\mathrm{TP}_k}{\mathrm{TP}_k + \mathrm{FP}_k},
\qquad
R_k = \frac{\mathrm{TP}_k}{\mathrm{TP}_k + \mathrm{FN}_k},
\qquad
F1_k = \frac{2\,P_k\,R_k}{P_k + R_k}
\label{eq:perclass}
\end{equation}

\begin{equation}
\text{Macro-}M = \frac{1}{K}\sum_{k=1}^{K} M_k,
\qquad M \in \{P, R, F1\}
\label{eq:macro}
\end{equation}

\begin{equation}
\text{Weighted-}M = \frac{1}{N}\sum_{k=1}^{K} n_k\,M_k,
\qquad M \in \{P, R, F1\}
\label{eq:wavg}
\end{equation}

\begin{equation}
\mathrm{Support}_k = n_k
\label{eq:support}
\end{equation}

These metrics are only reported on the Macro version, since the four classes have the same quantity of samples (150 images), and weighted and macro metrics represent the same value. All metrics were computed on the held-out test set. Additionally, a confusion matrix was used as a supporting metric, calculated for all models but not reported for each one.

\subsection{Experimental setup}
All of the results were generated under the same environmental setup: a computer equipped with an Intel Core i9 Central Processing Unit (CPU) and an NVIDIA GeForce RTX 5060 Graphics Processing Unit (GPU). All experiments were performed using Python 3.12.10, with libraries such as PyTorch and torchvision for model definition, pretrained weight loading, and training; scikit-learn for computing evaluation metrics and confusion matrices; and NumPy for deterministic data partitioning. A fixed random seed of 42 was used throughout, so that the class-balanced split described in Section~\ref{subsec:dataset} is reproducible across all ten architectures.


\section{Experimental Results}
\label{sec:results}

After training all the CNNs under the two-stage transfer-learning and fine-tuning pipeline, they were evaluated through the same evaluation protocol using a held-out test split of 150 images per class (600 images in total). The results of benchmarking these different architectures are shown in Table \ref{tab:results}. 

\begin{table}[H]
\caption{Test-set performance of the ten evaluated architectures under the
same protocol, ordered by test accuracy. Precision, recall, and F1 are
macro-averaged.}
\label{tab:results}
\begin{center}
\renewcommand{\arraystretch}{1.3}%
\begin{tabular}{lcccccc}
\hline
\rule[-1ex]{0pt}{3.5ex} Architecture & Val.\ acc. & Test acc. & Precision & Recall & F1 & Errors \\
\hline
\rule[-1ex]{0pt}{3.5ex} VGG16           & 0.9637 & 0.9533 & 0.9536 & 0.9533 & 0.9531 & 28 \\
\rule[-1ex]{0pt}{3.5ex} ResNet18        & 0.9375 & 0.9400 & 0.9395 & 0.9400 & 0.9394 & 36 \\
\rule[-1ex]{0pt}{3.5ex} ResNet34        & 0.9525 & 0.9400 & 0.9397 & 0.9400 & 0.9397 & 36 \\
\rule[-1ex]{0pt}{3.5ex} ResNet50        & 0.9363 & 0.9333 & 0.9343 & 0.9333 & 0.9331 & 40 \\
\rule[-1ex]{0pt}{3.5ex} ResNeXt50-32x4d & 0.9388 & 0.9250 & 0.9285 & 0.9250 & 0.9258 & 45 \\
\rule[-1ex]{0pt}{3.5ex} ResNet101       & 0.9388 & 0.9217 & 0.9259 & 0.9217 & 0.9222 & 47 \\
\rule[-1ex]{0pt}{3.5ex} DenseNet121     & 0.9300 & 0.9167 & 0.9188 & 0.9167 & 0.9169 & 50 \\
\rule[-1ex]{0pt}{3.5ex} EfficientNet-B0 & 0.8963 & 0.9033 & 0.9065 & 0.9033 & 0.9034 & 58 \\
\rule[-1ex]{0pt}{3.5ex} DenseNet169     & 0.9350 & 0.9000 & 0.9039 & 0.9000 & 0.9003 & 60 \\
\rule[-1ex]{0pt}{3.5ex} MobileNetV3-L   & 0.9163 & 0.8817 & 0.8882 & 0.8817 & 0.8827 & 71 \\
\hline
\end{tabular}
\end{center}
\end{table}

The best-performing architecture was VGG16 on all four metrics (Accuracy: 0.9533, Precision: 0.9536, Recall: 0.9533, F1: 0.9531), committing 28 errors. The best architectures from the ResNet family are ResNet18 (0.9400) and ResNet34 (0.9400), outperforming the larger models in the same family: ResNet50 (0.9333), ResNeXt50-32x4d (0.9250), and ResNet101 (0.9217). Additionally, the worst-performing CNNs were among the densely connected and efficient or mobile-optimized architectures, with the worst performance reached by MobileNetV3-L (0.8817), which nonetheless remains a competitive result, given that this is the smallest encoder (4.2M parameters). The total difference between the best-performing model and the worst is 0.072 points, which is not a substantial gap. This suggests the two-stage training protocol worked and that the first 5 transfer learning epochs helped the model stabilize before the 20 full fine-tuning epochs. 

Additional findings indicate that no model appears to suffer severe overfitting. Most models showed a validation-to-test difference below 0.02, suggesting they generalize well on unseen data. The exceptions are DenseNet169 and MobileNetV3-L, both above 0.03, which are also the two worst-performing models on the held-out test split. Additionally, VGG16 did not overfit regardless of being the largest model (134.3M parameters).

\subsection{Depth does not improve accuracy within architecture families}
\label{subsec:depth}
An interesting finding made visible after evaluating the ten different CNNs is that performance degrades consistently on larger architectures within the same family or design. For example, this is most notable on the ResNet family, where ResNet18 (11.3M parameters) achieves a 0.9400 test accuracy, while ResNet101 (43.0M parameters) achieves 0.9217 test accuracy. ResNet101 has 3.8 times more parameters than ResNet18 and is still outperformed by the smaller version. These examples are present across all ten CNNs benchmarked and are presented in Table \ref{tab:depth}.

\begin{table}[H]
\caption{Effect of depth within architecture families. Within each residual and
densely-connected family, increasing depth consistently reduced test accuracy.}
\label{tab:depth}
\begin{center}
\renewcommand{\arraystretch}{1.2}%
\begin{tabular}{llrrrr}
\hline
\rule[-1ex]{0pt}{3.5ex} Family & Variant & Total params & Val.\ acc. & Test acc. & Gap (pp) \\
\hline
\rule[-1ex]{0pt}{3.5ex} Residual   & ResNet18   & 11.3\,M & 0.9375 & 0.9400 & $-0.25$ \\
\rule[-1ex]{0pt}{3.5ex}            & ResNet34   & 21.4\,M & 0.9525 & 0.9400 & $+1.25$ \\
\rule[-1ex]{0pt}{3.5ex}            & ResNet50   & 24.0\,M & 0.9363 & 0.9333 & $+0.30$ \\
\rule[-1ex]{0pt}{3.5ex}            & ResNet101  & 43.0\,M & 0.9388 & 0.9217 & $+1.71$ \\
\hline
\rule[-1ex]{0pt}{3.5ex} Densely conn. & DenseNet121 &  7.2\,M & 0.9300 & 0.9167 & $+1.33$ \\
\rule[-1ex]{0pt}{3.5ex}               & DenseNet169 & 12.9\,M & 0.9350 & 0.9000 & $+3.50$ \\
\hline
\rule[-1ex]{0pt}{3.5ex} Residual vs.\ grouped & ResNet50 & 24.0\,M & 0.9363 & 0.9333 & $+0.30$ \\
\rule[-1ex]{0pt}{3.5ex}                       & ResNeXt50-32x4d & 23.5\,M & 0.9388 & 0.9250 & $+1.38$ \\
\hline
\end{tabular}
\end{center}
\end{table}

There is no ground-truth reason for these larger models to degrade in performance in a consistent manner on all design families. An attributable factor is the mismatch between model capacity and the available data quantity from this classification task, because larger models can learn more representations and are trained on ImageNet, spanning a million images across a thousand classes, but when, during training, they are introduced to only 800 images per class, these models can overfit or memorize the training set to some extent. This can be seen in Table \ref{tab:depth}, where test accuracy decreases monotonically with depth in every family, and the largest variant of each family shows a bigger validation-to-test gap than the smallest model. Nonetheless, this is attributed to the gradual degradation of performance between smaller and larger models. The observed gaps indicate a mild tendency toward overfitting that grows with capacity, not severe overfitting, which was not observed in any of the evaluated models.  

Building on this, VGG16 is the largest model in terms of parameter count, yet the top performer. This is because 89\% of the model parameters are in the classifier, which is pretrained and is not the main component responsible for extracting key features from the images. Consequently, the total parameter count on the backbone of the CNN is 14,713,536 (14M), similar to the ResNet18 architecture (11.3M). Consistent with this, VGG16 has one of the smallest validation-to-test gaps (+1.04\, pp) despite its size.

\subsection{The difficulty in classifying Non-Demented and Very Mild Demented samples}
\label{subsec:ndvsvmd}
The second key finding in this study was the notable difficulty all models had in classifying Non-Demented and Very Mild Demented samples. This is due to the structural similarity between these two categories; there is not an aggressive change from a non-demented brain to a very mild demented brain, whereas the difference from a non-demented brain to a moderate demented brain is far more pronounced. Additionally, this study works with 2D slices instead of a 3D image, where these distinctions could be more apparent. 

\begin{figure}[H]
\centering
\includegraphics[width=0.85\linewidth]{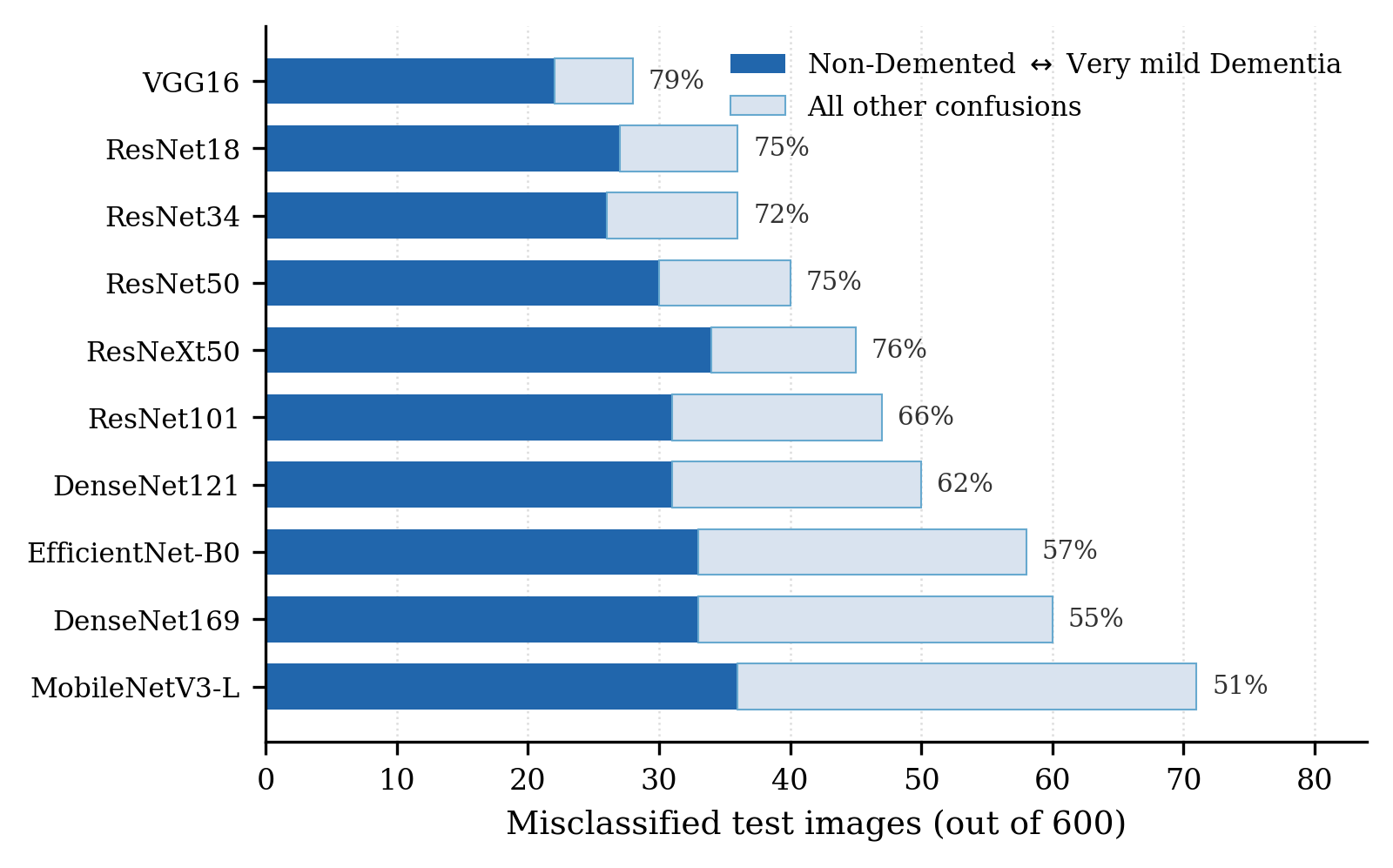}
\caption{Error decomposition across all ten architectures, ordered by test
accuracy. Between 51\% and 79\% of every model's misclassifications fall on the
single Non-Demented\,/\,Very Mild Dementia boundary.}
\label{fig:errordecomp}
\end{figure}

The distribution of these errors on the non-demented (ND) and very mild demented (VMD) samples is illustrated in Fig.~\ref{fig:errordecomp}. As shown in this figure, the distribution of mistakes across the ND and VMD classes remains similar across the 10 models. This suggests all architectures struggle with these two classes, and when the error increases, it usually increases in the other 2 classes, but the distribution of misclassifications on ND and VMD classes remains similar.   

\begin{figure}[H]
\centering
\includegraphics[width=\linewidth]{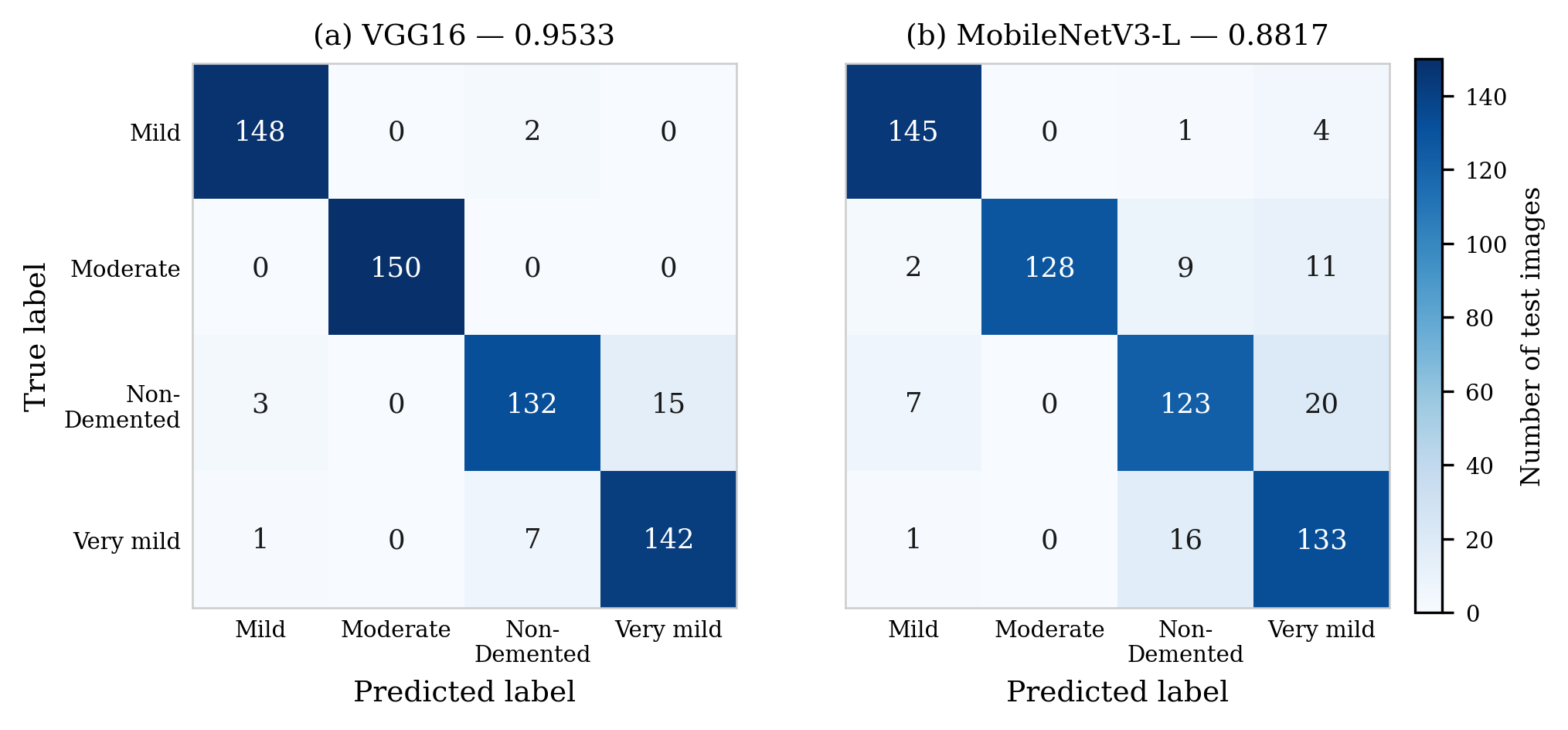}
\caption{Confusion matrices on the held-out test set for the best (VGG16) and
worst (MobileNetV3-L) models.}
\label{fig:confusion}
\end{figure}

To provide a good example, the two confusion matrices of the best- and worst-performing models (VGG16 and MobileNetV3-L) are shown in Fig.~\ref{fig:confusion}. The confusion matrices show that both models struggle most with the Non-Demented and Very Mild classes, comprising 78.6\% of the errors on VGG16 (22 out of 28 errors) and 50.7\% of the errors on MobileNetV3-L (36 out of 71 errors). This also suggests that as the model improves, it reduces other errors but continues to struggle with classifying the ND and VMD classes. 

\subsection{Performance versus computational cost}
\begin{figure}[H]
\centering
\includegraphics[width=0.75\linewidth]{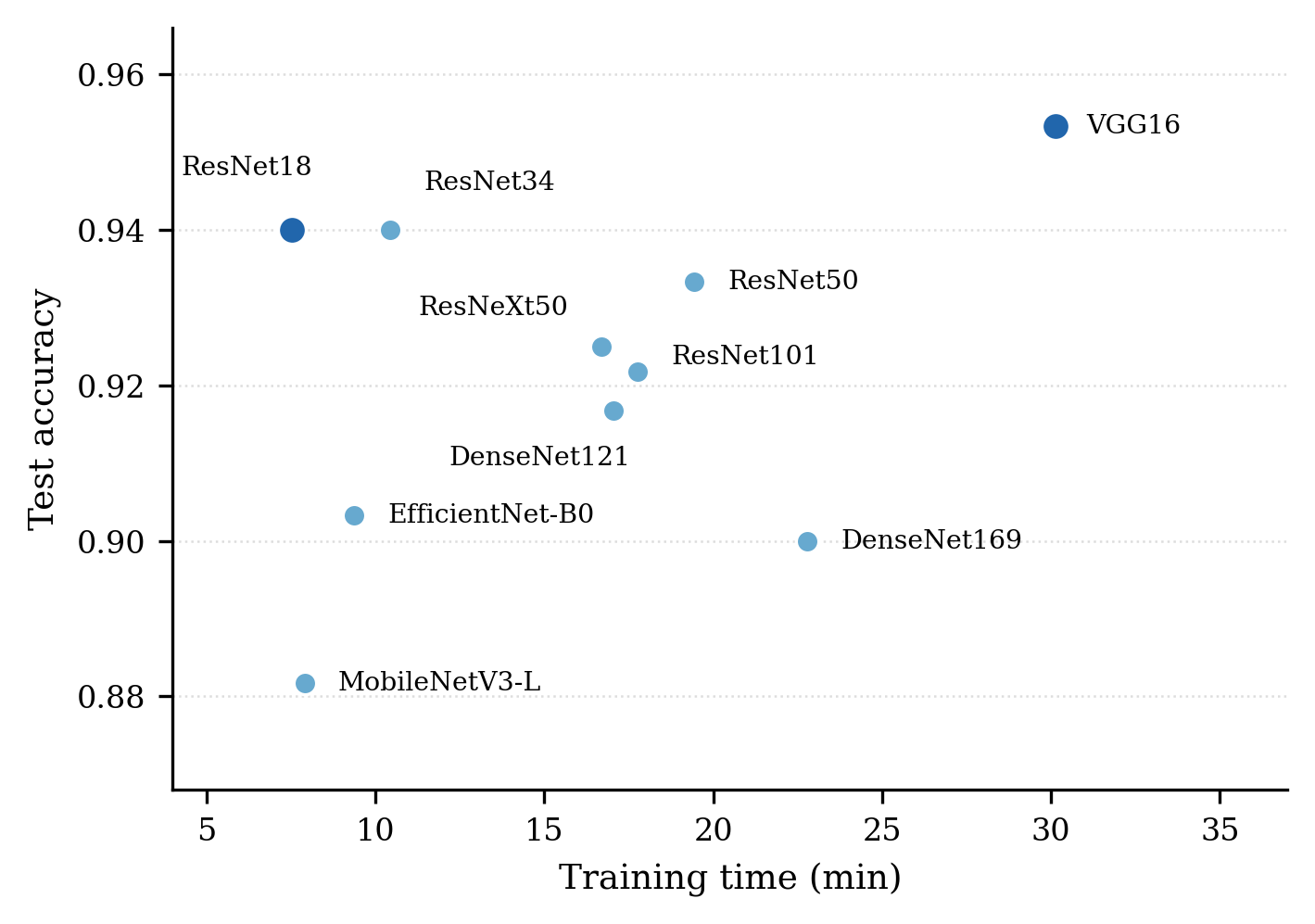}
\caption{Test accuracy versus training time. ResNet18 reaches 98.6\% of VGG16's accuracy in one quarter of the training time, while MobileNetV3-L, designed for efficiency, takes as long as ResNet18 yet scores 0.0583 lower on accuracy.}
\label{fig:acctime}
\end{figure}

A further analysis of how cost-effective these architectures are is illustrated in Fig.~\ref{fig:acctime}. The main finding is that, even though VGG16 achieved the highest performance, it also required the most computational time to train, taking 30 minutes to complete the 25 training epochs. In contrast, ResNet18 took only 7 minutes to train the same 25 epochs and achieved a 0.9400 accuracy score. Compared to the 0.9533 score of the VGG16, ResNet18 is a more cost-effective solution for Alzheimer's disease classification, given that the gap between these two models is only 0.0133 in accuracy.

Another finding was made when comparing ResNet18 and MobileNetV3-L; the latter CNN is optimized for mobile use with short inference and training times. This is true because both CNNs took 7 minutes to train, but MobileNet was the worst-performing CNN, and ResNet18 was the second best-performing CNN. Therefore, even though they are the two fastest architectures, the performance tradeoff is notable on ResNet18. These two findings position ResNet18 as a great alternative to VGG16 due to its training time, which would be a great advantage if we were to train on the entire OASIS dataset, comprising 86,437 images. 

Additionally, ResNet50 took more time to train (19 minutes) than ResNet18 (7 minutes) and had inferior performance (0.9333 vs. 0.9400), reinforcing the idea discussed in Section \ref{subsec:depth} that deeper models do not necessarily contribute beyond computational cost. The same pattern is visible with DenseNet169 and ResNet101. These results should be interpreted with the understanding that the study used an RTX 5060 GPU to generate the results and serves as a model comparison; other studies may report different times due to different experimental environments.

\subsection{Training discussion}
VGG16 is the CNN architecture with the highest performance across the four metrics among the ten architectures. As previously discussed in section \ref{subsec:ndvsvmd}, this model also struggles in differentiating samples from the Non-Demented class and the Very Mild Dementia class. This is evident in the individual performance table of the VGG16 model presented in Table \ref{tab:perclass}, where the non-demented and very mild dementia classes have the lowest precision, recall, and F1 scores. Another tendency visible only in this table is the preference to classify patients as Alzheimer positives rather than classify them into a non-demented class when the model is uncertain about the classification. This is shown in the recall metric across the Alzheimer-positive classes: Mild Dementia: 0.9867, Moderate Dementia: 1.0000, and Very Mild Dementia: 0.9467, against a 0.8800 recall value for the Non-demented class. This specific behavior is beneficial in this setting, where sending a person to a doctor for further clinical review is preferred rather than classifying him as non-demented and delaying a doctor's diagnosis.   

\begin{table}[H]
\caption{Per-class test performance of the best model (VGG16).}
\label{tab:perclass}
\begin{center}
\begin{tabular}{lcccc}
\hline
\rule[-1ex]{0pt}{3.5ex} Class & Precision & Recall & F1 & Support \\
\hline
\rule[-1ex]{0pt}{3.5ex} Mild Dementia      & 0.9737 & 0.9867 & 0.9801 & 150 \\
\rule[-1ex]{0pt}{3.5ex} Moderate Dementia  & 1.0000 & 1.0000 & 1.0000 & 150 \\
\rule[-1ex]{0pt}{3.5ex} Non-Demented       & 0.9362 & 0.8800 & 0.9072 & 150 \\
\rule[-1ex]{0pt}{3.5ex} Very Mild Dementia & 0.9045 & 0.9467 & 0.9251 & 150 \\
\hline
\rule[-1ex]{0pt}{3.5ex} Macro avg          & 0.9536 & 0.9533 & 0.9531 & 600 \\
\hline
\end{tabular}
\end{center}
\end{table}

\begin{figure}[H]
\centering
\includegraphics[width=\linewidth]{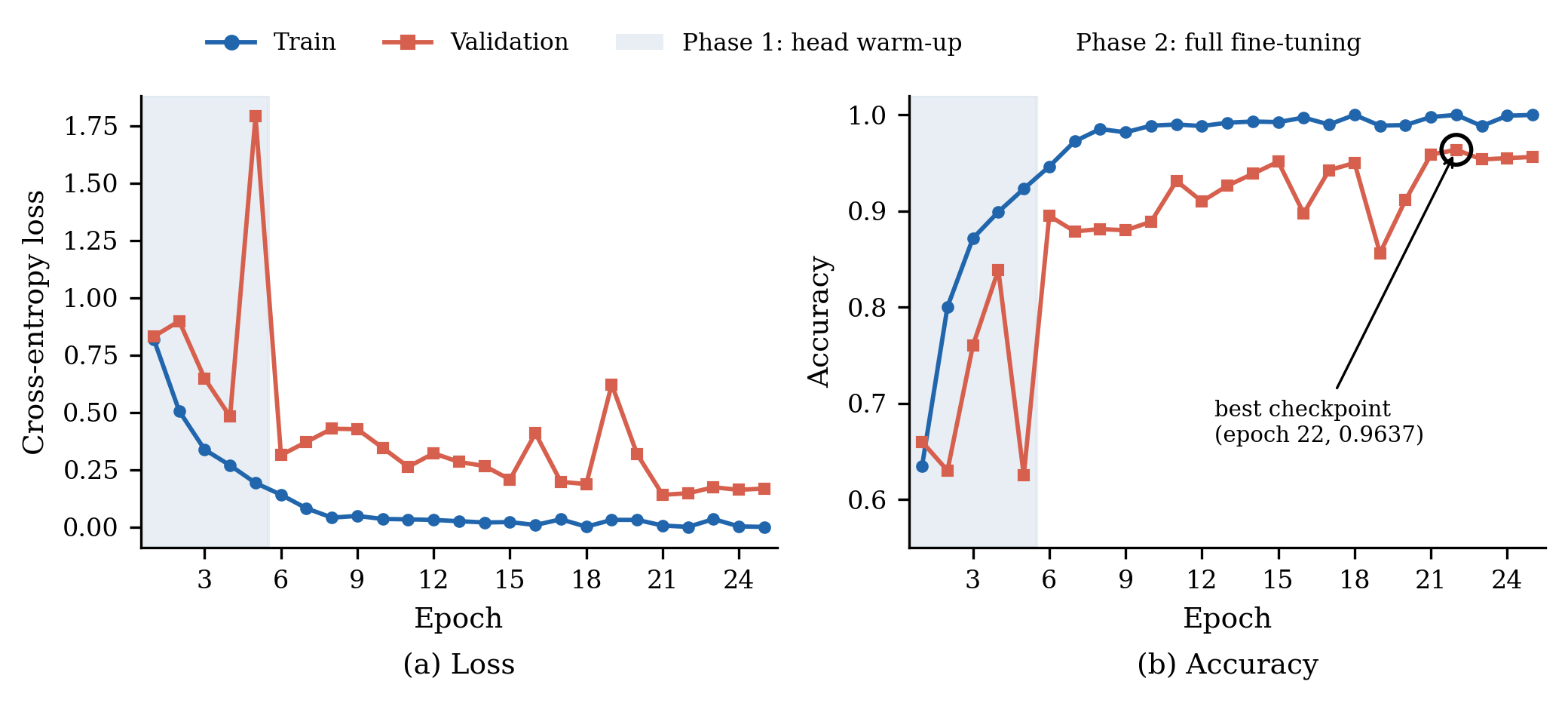}
\caption{Two-phase training curves of VGG16 across 25 epochs. Panel~(a) shows
cross-entropy loss and panel~(b) shows accuracy, each reporting the training and
validation sets. The shaded region corresponds to Phase~1 (transfer learning), and the remainder to Phase~2 (full fine-tuning). The circled marker in panel~(b) indicates the checkpoint selected for evaluation, chosen by the highest
validation accuracy.}
\label{fig:vggcurve}
\end{figure}

The two-phase training curves of VGG16 are presented in Fig.~\ref{fig:vggcurve}. During phase 1, with the backbone frozen, validation loss is unstable and peaks promptly around epoch 5 alongside accuracy. The transition to full fine-tuning in phase 2 immediately stabilizes both validation loss and accuracy, which supports the choice of the two-stage training protocol. A notable feature is that validation accuracy continues to improve until epoch 22, where it achieves the best checkpoint with 0.9637 accuracy, even after training accuracy has already reached 1.0000. This suggests the model keeps generalizing after it has memorized the training set, which is why the best checkpoint is selected by validation accuracy rather than using the score at the final epoch.

\subsection{Limitations}
Some limitations should be considered when interpreting these results and are described as follows:

\begin{itemize}
  \item \textbf{Slice-level partitioning.} The dataset was split at the level of
  individual 2D slices rather than by subject, so slices from the same scan may appear in both the training and test sets. The relative ranking across architectures produced by the benchmark remains valid, since all models were evaluated on the identical partition.

  \item \textbf{Non-uniform classification head.} The head is not matched across
  all ten architectures: VGG16 retains its pretrained classifier while the
  others use a randomly-initialized head.

  \item \textbf{Single cohort.} All images used for evaluation come from a single dataset (OASIS).
  Generalization under domain shift, to images acquired with different scanners and
  acquisition protocols remains untested.
\end{itemize}


\section{Conclusions \& Future Work}
\label{sec:conclusions}
This work introduces a benchmark of ten different CNN architectures evaluated under the same protocol, consisting of a held-out test split only used to report final results. Training was performed using a proposed two-stage transfer-learning and fine-tuning pipeline, whose objective is to stabilize head adaptation before proceeding to full fine-tuning, potentially avoiding overfitting given the limited size of the training split (3200 images) and the large quantity of parameters in some of the evaluated models. VGG16 was the best-performing architecture, achieving a test accuracy score of 0.9533 on the held-out test split. Additionally, ResNet18 established itself as a strong cost-effective alternative (0.9400 accuracy), taking 22 fewer minutes to train (7.5 vs. 30.1 minutes) while trading off only 0.0133 in accuracy. 

Two main findings were visible consistently across the benchmark. First, within each model family, larger variants degraded performance rather than improved test accuracy. This is attributable to a mismatch between model capacity and the limited training data. Second, across all models evaluated, most errors are concentrated on the Non-Demented and Very Mild Dementia classes. This is due to structural similarity in the brain between these two categories, where an aggressive change is not notable. Future work could focus on evaluating these CNNs under a subject-level protocol to eliminate slice-level data leakage, extending the benchmark with Vision Transformers (ViT), or evaluating on an external held-out dataset to quantify performance under domain shift.

\acknowledgments        
 
This work was supported by the Coordinaci\'on Institucional de Investigaci\'on of CETYS Universidad.

\bibliography{report} 
\bibliographystyle{spiebib} 

\end{document}